# On the ocean beach - why elliptic pebbles do not become spherical

Klaus Winzer* and Gerhard C. Hegerfeldt**

*I. Physikalisches Institut and **Institut für Theoretische Physik, Universität Göttingen, Friedrich-Hund-Platz 1, D-37077 Göttingen, Germany*

**Abstract**

Among pebbles strewn across a sandy ocean beach one can find relatively many with a nearly perfect elliptical (ellipsoidal) shape, and one wonders how this shape was attained and whether, during abrasion, the pebbles would remain elliptical or eventually become spherical. Mainly the latter question was addressed in a previous publication which identified frictional sliding and rotation of an elliptic pebble as main abrasion processes in the surf waves. In particular, it was predicted that the ellipticity $\varepsilon = \{1 - b^2/a^2\}^{1/2}$, ($a > b$ principal ellipse axes) converges to a common equilibrium value for elliptic-like pebbles. Unfortunately, the derivation was based on an invalid force expression and a dimensionally unsuitable curvature. In this paper, not only force and curvature but also the contact duration with the sand surface during rotations is taken into account by fairly simple physical arguments, and it is shown that elliptic pebbles neither approach the same ellipticity and nor become more spherical nor more disk-like but rather that the ellipticity $\varepsilon$ increases.

## 1. Introduction

The first explanation for the smooth, rounded shape of the pebbles was probably given by Aristotle [1]. He proposed that the abrasion is more efficient at regions far from the centre of the pebble where greater impulses can be more readily delivered. Aristotle himself claimed that spherical shapes are dominant. In the past, some field and laboratory studies on pebble shape have been published [2-7], with the mutual abrasion of pebbles by collisions in mind, like in a riverbed or on a pebble beach. Also, a number of mathematical models for the evolution of pebbles were published [8-12]. References [11] and [12], in particular, do not take into account the frictional abrasion process due to rotation around the axis of greatest moment of inertia. In refs. [13] and [14] the importance of this process was investigated. Indeed, any observer will immediately notice this rolling motion of a pebble in the water backflow on the beach. Observable evidence for its importance was pointed out in refs. [4] and [14].

The abrasion of an elliptic-like pebble on a sandy beach readily lends itself to a detailed physical analysis and model building, as shown in ref. [13] on which the present paper builds. About 1500 ellipsoidal pebbles were collected by one of the authors (K.W.) on flat sandy beaches. In particular, on the south beach of Heligoland, one can find elliptical pebbles formed from brick fragments produced during the blasting of military buildings after the Second World War in 1947. The time span of about 70 years allows an estimation of the formation time for an elliptical pebble of brick. With the knowledge of the abrasion hardness and the densities one can estimate the formation time for an elliptical pebble of the much harder mineral basalt, leading to times of the order of 1000 years [14].

*E-mail: kwinzer1@gwdg.de

Unfortunately, in ref. [13] an invalid force expression and a dimensionally unsuitable curvature expression were used, also contact durations were not taken into account; as a consequence, in particular the claimed convergence of the ellipticity $\varepsilon$ to a common equilibrium value due to rotation around the c-axis does not hold, rather the ellipticity $\varepsilon$ will increase, as will be shown further below.

The plan of the paper is as follows. In sect. 2 basic properties of the collection of elliptical pebbles are described. In sect. 3 observations of the motion of elliptical pebbles are used to derive a model for their mathematical description. In sect. 4.1 this model is used to derive the correct force with which an elliptic pebble acts on the sandy beach. Since not the force but force per area is relevant for the abrasion the contact area is needed. This area becomes smaller the larger the curvature is. Therefore, in sect. 4.2, we make the ansatz that the contact area is proportional to the inverse Gaussian curvature at the contact point. The Gaussian curvature has the correct dimension of an inverse area, while the so-called mean curvature, used in ref. [13], has the dimension of an inverse length. In the surf, the abrasion at a given pebble point depends on the contact time of this point with the wet sand. By simple physical arguments we derive an expression for this contact time which, up to a constant, holds for any elliptic pebble. These three ingredients are used to derive differential equations for the temporal development of the abrasions of the a- and b-axis for rotations around the c-axis. In sect. 4.3 this is carried over to rotations around the a-axis. From these equations it is shown analytically that the ellipticity $\varepsilon$ of a pebble increases in time. Hopping shows up as a singularity in the differential equations. Eventually, when the b/a ratio decreases roughly below 0.5, increased hopping and tumbling will destroy the ellipsoidal-like shape of a pebble. In sect. 5 the results are discussed.

**2. Empirical basis: A collection of 1500 elliptic-like pebbles**

Over the course of about ten years, one of the authors (K.W.) collected about 1500 nearly ellipsoidal pebbles on flat sandy beaches, with low density of pebbles and thus negligible collisions between them. Only pebbles appearing outwardly isotropic were included. They were collected mostly on the beaches of the Canary Islands, the Cap Verde Islands and along the Turkish south coast between Alanya and Side. A smaller number was collected on the Baleares, Madeira, Porto Santo and on the North Sea island of Heligoland. Some elliptical pebbles of different sizes and from different beaches are shown in fig.1. The lengths of the a-axis of the collected pebbles vary from 0.4 to 8 cm and their masses from 0.2 to 2000 g. The most common values for the a-axis are about 2 cm and for the mass 40 g.

The Cap Verdes and the Canaries are of volcanic origin, with still partially active volcanos. Therefore, these pebbles consist of dark grey to black igneous basalt. The average density $\rho$ of these basalt pebbles is about 3.6 g cm$^{-3}$ and their abrasive hardness (according to Rosiwal [16]) is $H_{R,bas} \cong 70$. The pebbles from the beaches of the southern Turkish coast mainly consist of light grey and yellow-brown marble (calcite, Ca[CO$_3$]). Due to the different contamination of the calcite with other minerals, the density varies in the range of 2.6 – 2.9 g cm$^{-3}$. The abrasive hardness of pure calcite is $H_{R,cal} = 4.5$, significantly lower than that of basalt. The pebbles collected on the beaches of the islands of the Baleares, of Madeira and Porto Santo and of Heligoland are mainly of sandstone, with densities $\rho$ of $2.2 - 2.6$ g cm$^{-3}$, but with greatly varying abrasive hardness.

Neither approximately spherical nor cigar-shaped ($a > b \approx c$) pebbles were found. Pebbles with oblate shape ($a \approx b > c$) are extremely rare. The percentage of elliptical pebbles relative to the total number of pebbles is only in the single digit range. It was argued [14] that this is due to the different initial forms of the edged rock fragments. Stones with an elongated shape always have an axis of largest moment of inertia, which for energetic reasons is the preferred axis of rotation. This allows the formation of the elliptic shape after a sufficiently long abrasion process. Stones more similar to a cube or tetrahedron have three axes of inertia whose moments of inertia differ very little. These stones become also smaller and rounder by rolling on the sandy beach, but not elliptical, because they do not have a preferred rotation axis. Since the break-off of an elongated polyhedron from a rock is much less likely than the break-off of a more compact fragment, only a small fraction of all pebbles have an elliptical shape.

In ref. [14] deformed ellipsoids (ovoids) with different curvatures on both ends of the longest axis (a-axis) were considered. It was shown that the curvatures on the a-axes of the ovoid are aligned with each other during the rotation around the shortest axis (c-axis), because the more strongly curved side of the ovoid was more strongly abraded. This alignment of the two curvatures occurred about three times faster than the decrease in the length of the a-axis. The elliptical shape of the pebble emerged much more quickly than the decrease in volume due to the shortening of the a-axis, providing a reason for the relatively large number of elliptical pebbles.

From the collected elliptical pebbles, frequency distributions as a functions of the axial ratios were obtained in ref. [14]. These distributions were shown to depend on the density of the pebble species. Since for the discussion below a constant density is needed we display in fig. 2 frequency distributions of 750 elliptical pebbles of the uniform mineral calcite as a function of the axial ratios $b/a$, $c/a$, and $c/b$. In a Gaussian fit, the maxima of the distributions are located at $b/a = 0.76$, $c/a = 0.398$, and $c/b = 0.534$, respectively. The relatively narrow frequency distributions for $b/a$ and $c/a$ point to a dominant grinding process in each case. The full width at half maximum (FWHM) for both distributions are nearly equal, with $FWHM_{b/a} = 0.205$ and $FWHM_{c/a} = 0.204$, respectively. Therefore, in the examples below we use 0.65, 0.76 and 0.87 as typical values for $b/a$.

### 3. Observations of pebble motion in the surf: The model

The following draws heavily on ref. [13]. There, different abrasion processes were identified: An incoming wave may lift up pebbles and when it slows down on the inclined sandy beach, first the larger, then the smaller pebbles fall to the ground and roll up the beach. When the water flows back to the sea with increasing speed three different scenarios may occur. (i) In a weak backflow, a pebble, with axes $a > b > c$ and lying on its flat side, slides seawards without rolling motion. The abrasion of the c-axis was shown to increase with decreasing $b/a$-ratio, and thus elongated pebbles are thinner on average than rounder pebbles - in good agreement with observation. (ii) A somewhat stronger backflow may cause the pebble to rotate around the longer a-axis since the potential energy for raising the pebble, $W_{pot,a} = m\,g\,(b - c)$, is smaller than for the c-axis. (iii) The most interesting case occurs for strong backflow. In a medium with friction, such as a water-sand mixture, only the rotation around the axis of the greatest moment of inertia is stable [15]. Therefore, the pebble quickly rises and performs a fairly stable rotation around the c-axis. In contrast to the brief rotation around the a-axis, often 10-20 rotations around the c-axis were observed and the pebble rolled several meters. Observations of rolling pebbles of the kind considered here (1 cm < a < 5 cm, 0.7 cm < b < 3 cm) show that they rotate with an essentially constant angular velocity and travel with a constant speed. This does not seem unreasonable when one considers the forces acting on a pebble, such as gravity, water, the sand, and friction. A torque is generated by the upward force exerted by the sand; by symmetry its net effect on the angular momentum over a single rotation vanishes. In addition, there is a torque due to friction during slipping. This latter torque points in the opposite direction of the former and so counteracts it, while the torque due to the velocity gradient of the water near the ground reinforces it. These combined forces and the resulting torques seem to lead to an effective translational velocity and to an effective angular velocity.

This rotation leads to a very effective abrasion process on a narrow surface strip perpendicular to the ab-plane. When the angular velocity becomes too large hopping sets in. Since the velocity of the water is greater than the translational velocity of the rotating pebble, water vortices are created in front of the pebble. When b/a is small, i.e. if the pebble is longish and more pencil-like, it will be more difficult for the wave to raise the pebble, by energetic reasons. Instead of a stable rotation around the c-axis the pebble will then more likely perform a chaotic tumbling motion which will not preserve ellipticity. This seems to be the reason why empirically no pebbles with $b/a < 0.5$ have been observed.

From these observations it follows that the erosion of an elliptic-like pebble proceeds via an intermittent sequence of sliding on the c-axis, rolling around the a-axis, and rolling around the c-axis. When the pebble becomes too small or when the rotation becomes too fast hopping sets in. A criterion for the onset of hopping will be given in eq. (21) below.

For rotations around the c and a axis these empirical observations suggest the following model for typical elliptic-like pebbles with $1 \text{ cm} < a < 5 \text{ cm}$.

1. These pebbles rotate with nearly constant angular velocity $\omega$ and the centre of mass has a nearly constant horizontal velocity component $v_{tr}$. As a reasonable connection we take $v_{tr} = \omega L/2\pi$ where $L$ is the circumference of the ellipse.

Typical observational values for a standard pebble with $a = 2$ cm and $b = 1.5$ cm are 3 Hz for the rotational frequency and 0.34 m/s for the horizontal velocity component. For smaller pebbles, with a ≤ 0.5 cm, the above will not apply.

2. This will lead to an up-and-down motion of the centre of mass and consequently to a time-dependent force between the pebble and on the sand layer.
3. As a further consequence, depending on the ellipticity $\varepsilon$, an elliptic-like pebble will experience a more or less pronounced slippage, i.e. in general the relative speed between the pebble surface and sand at the contact point will not be zero.

When the pebble velocity becomes too large then hopping sets in and the above model is no longer appropriate. Similarly when the mass becomes too small.

These ingredients will be exploited in the next section, where in particular the force and the slippage will be explicitly calculated.

## 4. Abrasion due to rolling

### 4.1. The relevant forces

The most effective mechanism of erosion apparently occurs for rolling and sliding. In contrast to sliding on the beach, in the case of rolling around the c-axis the force is not constant because the centre-of-mass of the ellipsoid performs an up-and-down movement. Figure 3 schematically shows the rolling of a pebble on the sand base, with $\varphi$ the angle between the a-axis and the radius $r$ from the centre-of-mass (CM) to the support point P$_S$.

To describe the dynamics of the pebble rotating around the c-axis, we calculate the vertical distance $r_{CM}(\varphi)$ between the supporting surface and the centre-of-mass (CM), as well as the vertical centre-of-mass velocity $v_{CM}(\varphi)$ and the corresponding acceleration $a_{CM}(\varphi)$. The distance to the centre-of-mass,

$$r_{CM}(\varphi) = r(\varphi) \sin(\alpha + \varphi), \qquad (1)$$

contains the angle $\alpha$ between the a-axis and the sandy ground as tangential plane. Using the well-known tangent equation for an ellipse one finds

$$\tan \alpha = \left(\frac{b}{a}\right)^2 (1/\tan \varphi). \qquad (2)$$

The radius $r(\varphi)$ from the center-of-mass to the supporting point P$_S$ is given by

$$\frac{r(\varphi)}{a} = \left(\frac{b}{a}\right) \frac{1}{\cos \varphi \sqrt{(b/a)^2 + \tan^2 \varphi}}. \qquad (3)$$

From the three relations (1 - 3), the distance $r_{CM}(\varphi)$ between P$_\perp$ and CM is obtained after a short calculation as

$$\frac{r_{CM}(\varphi)}{a} = \left(\frac{b}{a}\right) \frac{\sqrt{(b/a)^2 + \tan^2 \varphi}}{\sqrt{(b/a)^4 + \tan^2 \varphi}}. \qquad (4)$$

Figure 4a shows $r_{CM}(\varphi)/a$ for three $b/a$-ratios with strong angular dependency around the a-axis and a weaker dependency around the b-axis. From fig. 3 the relation
$$\tan \alpha \cdot \tan(\omega t) = 1 \qquad (5)$$

between the tangent angle $\alpha$ and $\omega t$ can be obtained. With eq. (2) this gives the following simple relation between the angle $\varphi$ and the angular velocity $\omega$ of the rotation of the elliptic pebble,

$$\tan \varphi = \left(\frac{b}{a}\right)^2 \tan(\omega t) \quad .$$
(6)

Thus, the angle $\varphi$ has a nontrivial time dependence which enters the vertical centre-of-mass velocities and accelerations. For the first time derivative of the angle $\varphi$ one obtains

$$\frac{d\varphi}{dt} = \left\{\left(\frac{b}{a}\right)^2 \cos^2\varphi + \left(\frac{a}{b}\right)^2 \sin^2\varphi\right\} \cdot \omega \ . \tag{7}$$

With eq. (7) one obtains for the vertical centre-of-mass velocity

$$\frac{v_{CM}(\varphi)}{a\,\omega} = \left\{\left(\frac{b}{a}\right)^2 \cos^2\varphi + \left(\frac{a}{b}\right)^2 \sin^2\varphi\right\} \cdot \frac{\tan\varphi}{\cos^2\varphi} \cdot \frac{(b/a)^3((b/a)^2-1)}{\sqrt{(b/a)^2+\tan^2\varphi}\cdot\left(\sqrt{(b/a)^4+\tan^2\varphi}\right)^3} \ . \tag{8}$$

Figure 4b shows $v_{CM}(\varphi)/a\omega$ for three $b/a$-ratios. For fixed $a$, the vertical centre-of-mass velocity strongly depends on the $b/a$-ratio: For slim pebbles with a small $b/a$-ratio the velocity is larger than the velocity of rounder pebbles with larger $b/a$-ratio. The magnitude of the centre-of-mass velocity has a maximum around $\varphi \cong 30°$. Therefore, for all pebbles the $\varphi$-dependence of the velocity is stronger near the a-axis than near the b-axis.

In the same way the vertical centre-of-mass acceleration $a_{CM}(\varphi)$ can be calculated. Figure 4c shows $a_{CM}(\varphi)/(a\omega^2)$ for the same $b/a$-ratios. Just as the velocity, the acceleration depends strongly on the $b/a$-ratios. For fixed $a$, the acceleration is larger for slim elliptic pebbles than for rounder ones. Although the slope of the velocity curves (fig. 4b) at the a-axis ($\varphi = 0$) is larger for all $b/a$ ratios than at the b-axis ($\varphi = \pi/2$), the magnitude of the acceleration at the a-axis is smaller than at the b-axis. This seeming contradiction is due to the complex relationship between the angle $\varphi$ and the angular velocity $\omega$ in eq. (6).

For the further discussion, the comparison of the vertical centre-of-mass accelerations at the a-axis ($\varphi = 0$) and at the b-axis ($\varphi = \pi/2$) is of importance. At the a-axis the vertical centre-of-mass acceleration is small and negative,

$$\frac{a_{CM}(\varphi=0)}{a\,\omega^2} = -\left(1 - \left(\frac{b}{a}\right)^2\right) \tag{9a}$$

while at the b-axis the acceleration is larger and positive,

$$\frac{a_{CM}(\varphi=\pi/2)}{a\,\omega^2} = +\frac{a}{b}\left(1 - \left(\frac{b}{a}\right)^2\right) \ . \tag{9b}$$

From this one obtains for the respective upward forces by the sand at the a- and b-axis

$$F_{s,a} = \frac{4\pi}{3}\rho\,abc\left\{g - a\,\omega^2\left(1 - \left(\frac{b}{a}\right)^2\right)\right\} , \tag{10a}$$

$$F_{s,b} = \frac{4\pi}{3}\rho\,abc\left\{g + a\,\omega^2\frac{a}{b}\left(1 - \left(\frac{b}{a}\right)^2\right)\right\} . \tag{10b}$$

These expressions will now be used to calculate the abrasions at the axes.

### 4.2 Abrasion of the a- and b-axis

The abrasion $\Delta a$ and $\Delta b$ of the a- and b-axis per revolution around the c-axis depends on the force per area and on the contact times, $\Delta t_a$ and $\Delta t_b$, with the sand at the respective axes. The contact time depends on the respective velocities, $a\omega$ and $b\omega$, and on the unknown contact length. It is natural to assume that the contact length is proportional to the respective curvature radii, $b^2/a$ and $a^2/b$, so that $a\omega\,\Delta t_a = \tilde{\gamma}\,b^2/a$ and $b\omega\,\Delta t_b = \tilde{\gamma}\,a^2/b$, where the unknown constant $\tilde{\gamma}$ takes into account the penetration depth of the pebble in the dense wet sand. Hence, with $\gamma \equiv \tilde{\gamma}/2\pi$ and $T$ the duration of one revolution, one has

$$\Delta t_a = \gamma\, (b/a)^2\, T \quad , \quad \Delta t_b = \gamma\, (a/b)^2\, T \ . \tag{11}$$

For the pressure, the unknown contact areas $A_a$ and $A_b$ are needed. We therefore assume that $A_a$ and $A_b$ are proportional to the inverse of the respective Gaussian curvature, which has the dimension of an area,

$$A_a = \gamma_A \frac{b^2}{a}\frac{c^2}{a} \quad , \quad A_b = \gamma_A \frac{a^2}{b}\frac{c^2}{b} \tag{12}$$

with a dimensionless constant $\gamma_A$. Thus we get for the abrasions

$$\Delta a = -\alpha_A \frac{F_{s,a}}{A_a}\, |v_{s,a}|\, \Delta t_a = -k_{ab}\frac{ab}{c}\left\{g - a\omega^2\left(1 - \left(\frac{b}{a}\right)^2\right)\right\}|v_{s,a}|\, T \quad , \tag{13a}$$

$$\Delta b = -\alpha_A \frac{F_{s,b}}{A_b}\, |v_{s,b}|\, \Delta t_b = -k_{ab}\frac{ab}{c}\left\{g + a\omega^2 \frac{a}{b}\left(1 - \left(\frac{b}{a}\right)^2\right)\right\}|v_{a,b}|\, T \tag{13b}$$

where $\alpha_A$ is the abrasion coefficient which strongly depends on the abrasion hardness of the mineral, where $k_{ab}$ contains all known and unknown constants, and where $v_{s,a}$ and $v_{s,b}$ are the yet to be determined slip velocities at the respective axes with which the elliptical pebble slips over the sand surface during the rotation around the c-axis. For a rotation around the c-axis with constant angular velocity the circumferential velocity at the contact point $P_s$ is

$$v(\varphi) = r(\varphi)\,\omega = \frac{b\,\omega}{\sqrt{(b/a)^2 \cos^2\varphi + \sin^2\varphi}} \ .$$

The horizontal (translational) velocity, $v_{tr}$, of the pebble is smaller than $v(\varphi)$ at the a-axis and larger at the b-axis. It is given by $v_{tr} = \omega\, L/2\pi$ where $L$ is the circumference of the ellipse. The ellipse circumference $L = L(a,b)$ is approximately given by $L \cong \pi\left[1.5(a+b) - \sqrt{ab}\right]$ which leads to $v_{tr} \cong \left[1.5\,(a+b) - \sqrt{ab}\right](\omega/2)$. The difference between $v(\varphi)$ and $v_{tr}$ is the slip velocity $v_s(\varphi)$, which is

$$v_{s,a}/(a\omega) \cong 1 - (1/2)\left[1.5\,(1 + b/a) - \sqrt{b/a}\right] \quad , \tag{14a}$$

$$v_{s,b}/(a\omega) \cong b/a - (1/2)\left[1.5\,(1 + b/a) - \sqrt{b/a}\right] \tag{14b}$$

at the a- and b-axis, respectively. Note that $v_{s,a} > 0$, $v_{s,b} < 0$ and $|v_{s,a}| < |v_{s,b}|$. Thus one obtains for the relative abrasions during one revolution

$$\frac{\Delta a}{a} = -k_{ab}|v_{s,a}|\frac{b}{c}\left\{g - a\,\omega^2\left(1 - \left(\frac{b}{a}\right)^2\right)\right\}\cdot T \ , \tag{15a}$$

$$\frac{\Delta b}{b} = -k_{ab}|v_{s,b}|\frac{a}{c}\left\{g + a\,\omega^2 \frac{a}{b}\left(1 - \left(\frac{b}{a}\right)^2\right)\right\}\cdot T \ . \tag{15b}$$

From this it follows that

$$\frac{\Delta b/b}{\Delta a/a} = \frac{a}{b} \cdot \frac{|v_{s,b}|}{|v_{s,a}|} \cdot \frac{g + a\,\omega^2 \frac{a}{b}\left(1-\left(\frac{b}{a}\right)^2\right)}{g - a\,\omega^2 \left(1-\left(\frac{b}{a}\right)^2\right)} > 1 \qquad (16)$$

since all quotients are greater than 1. Hence $|\Delta b|/b > |\Delta a|/a$ so that the relative abrasion at the b-axis is stronger than at the a-axis. Moreover, cancelling $a$ and $b$ on both sides of eq. (16), one has also $\Delta b/\Delta a > 1$. Hence the ellipticity $\varepsilon = \{1 - b^2/a^2\}^{1/2}$ increases during rotations around the c-axis, and there is no equilibrium, contrary to the claim in ref. [13]. Note that in eq. (16) the time-dependent $c$ has dropped out.

The dependence of $a$ on time will be explored further below. If instead of time one takes $a$ as variable - decreasing $a$ corresponding to increasing time - eq. (16) becomes a differential equation,

$$\frac{db(a)}{da} = \frac{|v_{s,b}|}{|v_{s,a}|} \cdot \frac{g + a\,\omega^2 \frac{a}{b}\left(1-\left(\frac{b}{a}\right)^2\right)}{g - a\,\omega^2 \left(1-\left(\frac{b}{a}\right)^2\right)} . \qquad (17)$$

Here $\omega$ will depend on $a$ and $b$. To find its dependence we assume that the translational velocity $v_{tr}$ of the pebble is roughly half of the backflow velocity $v_w$, i.e. $\omega = 2\pi\, v_{tr}/L(a,b)$, so that with $L \cong 2\pi\sqrt{ab}$ one obtains

$$a\,\omega^2 \cong v_w^2/4\,b. \qquad (18)$$

With an average backflow velocity of $v_w = 1$ m/s one obtains for a standard pebble ($a = 2\ cm, b = 1.5\ cm$) a rotation frequency of $\nu_o = 4.6$ Hz .

For given initial values of $a$ and $b$ we have solved eq. (17) numerically with the exact form of $\omega$. In fig. 5a, different curves are displayed which start at $a_o = 3$ cm with various $b$ values. Several features strike the eye. First, for larger $(b/a)_o$ ratio the curves start out almost linearly, with slope close to 1. Second, for decreasing $a$ the slope becomes steeper, the more so the smaller $b$ is. When the denominator in eq. (17) vanishes the numerical solution becomes singular at lower $a$ values (not shown in the figure). Taken literally, the longest principal axis $a$ would stop abrading whereas $b$ would continue to abrade. At this point hopping sets in and the model no longer applies. For the respective curves this occurs, from left to right, for $a = $ 1.24, 1.57, 1.80, 2.02, 2.23, 2.44, and 2.65 cm. This is also apparent in the corresponding curves for $b(a)/a$ in fig 5b. For decreasing $a$ one sees that $b$ decreases much faster than $a$. For a more disk-like pebble this means that during rotations around the c-axis the ablation of the b-axis will only be slightly larger than that of the a-axis. Then, after a long time, when $b$ has become appreciably smaller than $a$, the abrasion of $b$ becomes much faster until hopping or other erosion processes like tumbling set in.

Returning to eq. (15a,b), we introduce a scaled time variable $\tau = k_{ab}\,t$ and put $\Delta \tau = k_{ab}\,T$ which is short compared to the overall abrasion process. Then eq. (15a) and eq. (15b) can be written as a set of coupled differential equations for $da/d\tau$ and $db/d\tau$,

$$\frac{da}{d\tau} = -|v_{s,a}|\,\frac{ab}{c}\left\{g - a\,\omega^2\left(1-\left(\frac{b}{a}\right)^2\right)\right\}, \qquad (19a)$$

$$\frac{db}{d\tau} = -|v_{s,b}| \frac{ab}{c} \left\{ g + a\,\omega^2 \frac{a}{b}\left(1 - \left(\frac{b}{a}\right)^2\right)\right\} \tag{19b}$$

which now, however, still involve the unknown $c = c(\tau)$. To eliminate $c$ we use the empirical relation $c = \lambda\sqrt{ab}$ with $\lambda \cong 0.46$ [13]. This is an average relation which in general does not hold for individual pebbles. Alternatively, one can use an average value for $c/a \cong 0.40$ as suggested by the empirical data. The results are consistent and differ very little. The coupled differential equations can be solved numerically for $a(\tau)$ and $b(\tau)$. Figure 6 displays $a(\tau)$ and $b(\tau)$ for initial value $a(0) = 3$ cm and $b(0)/a(0) = 0.65, 0.76, 0.87$. The temporal decrease of $a$ and $b$ is almost linear, with slightly different slopes for $a$ and $b$, until a singularity is approached where hopping sets in. For this reason, the dependence of $b(a)$ on $a$ in eq. (17) mirrors the dependence on $t$. Again one sees a similar behaviour as in fig. 5a: for smaller $b/a$ ratio $b(\tau)$ decreases very fast until it diverges. The corresponding curves for $b(\tau)/a(\tau)$ (not shown) resemble those in fig. 5b.

In fig. 7 the collected calcite pebbles are displayed according to their $a$ and $b$ values. They are localized in a narrow band between two straight lines, namely $b = a$ and $b = a/2$. The line in between is a linear fit with slope 0.715 and interception 0.061. Superimposed is the development the pebbles would take by subsequent abrasion due to rotation around the c-axis. Depending on the initial $b/a$ -ratio they would rapidly approach the $b/a$ ratio ½ ; since elliptic-like pebbles of $b/a$ ratio ½ are not observed, as discussed in sect. 3, the model predictions seem to contradict empirical observations at this point and other abrasion processes take over. This might indicate a diminishing number of elliptical pebbles for diminishing value of $a$, contrary to the empirical evidence. Therefore, to preserve the dynamical equilibrium of the set of elliptical pebbles, new ones must appear which are created from non-elliptical fragments as discussed in sect. 2 and which replenish the set of previously existing pebbles.

### 4.3 Abrasion by rolling about the a-axis

From observations on the beach one concludes that the durations of rotations around the a-axis are in general much shorter than those around the c-axis, as are the corresponding abrasions. The results of the previous section apply directly to the present case, with $b$ taking the role of $a$ and $c$ that of $b$. This implies that, during rotations around the a-axis, the abrasion at the shorter c-axis is stronger than at the longer b-axis. The abrasion of the b-axis during this process adds to the abrasion of $b$ during rotations around the c-axis.

### 5. Discussion

The results of the preceding sections show that elliptical pebbles will become more elliptical by the above two types of grinding during water backflow until other mechanisms like hopping or tumbling set in. It is noteworthy that eqs. (16) and (17) remain qualitatively similar if one replaces the Gaussian curvature by the square (for dimensional reasons) of the mean curvature. This shows a certain robustness of our approach.

There is clear experimental evidence for rolling as a dominant effect for the shape of the ab-plane. As noted in the work of Rayleigh [4], the ab-plane usually shows slight deviations from the ideal elliptical shape. These deviations are due to the fact that the slip velocity in eqs. (14a) and (14b) is positive on the a-axis, but negative on the b-axis. During one revolution around the c-axis, the sign of the slip velocity changes four times. At these angles $\varphi(v_s = 0)$, the pebble will be less abraded because there is no slip. Therefore, in these areas one has $\Delta r = r(\varphi) - r_E(\varphi) > 0$. The amount of $\Delta r(\varphi)/r_E(\varphi) \cong 0.02 - 0.05$ is small, but it can be observed on all collected elliptical pebbles [14].

Figure 7 shows that almost no elliptical pebbles with b/a-ratios of less than about 0.5 are found. One reason for this is the dependence of the rotational energy on the axis ratios. For the rotation around the c-axis the rotational energy is given by

$$W_{rot,c} = \frac{I_c}{2} \cdot \omega^2 = \frac{m}{10} \cdot ab \cdot \left(\frac{a}{b} + \frac{b}{a}\right) \cdot \omega^2 \tag{20}$$

At constant $m$ and $\omega$, the rotational energy is lowest for $b = a$ and increases for $b < a$. The probability of the rotation around the c-axis therefore decreases for smaller $b/a$ values.

The $b/a$-ratio is determined by three erosion processes, which can be clearly separated depending on the area $A_{ab} = \pi\, ab$ of the ab-plane.

(i) For large and therefore heavy pebbles, the rotation mainly takes place around the a-axis, because the potential energy for raising a pebble, $W_{pot,a} = m\, g\, (b - c)$, is lowest for the rotation around the longest axis. Here only the b- and c-axes are eroded. Hence one expects a small $b/a$-ratio for pebbles with large $A_{ab}$.

(ii) In the case of smaller pebbles, due to their lower mass, the kinetic energy of the water is sufficient to provide the potential energy for raising the pebble to rotate around the c-axis. This rotation mode is energetically preferred because, at given angular momentum $L_\omega$ and moment of inertia $I_c = (m/5) \cdot (a^2 + b^2)$, the rotational energy $W_{rot,c} = L_\omega^2/2I_c$ is lowest. For medium-sized elliptical pebbles, the $b/a$-ratio should decrease as $A_{ab}$ decreases.

(iii) In water backflow, very small elliptical pebbles will have a tendency to hopping and tumbling, the more so the lighter they are. This is related to the fact that the angular momentum around the c-axis, $L_\omega = I_c \omega_c \propto a^4$, becomes very small [13], so that a stable rotation around the c-axis is no longer possible. For given small $A_{ab}$ hopping and tumbling is more likely for pebbles with small b/a-ratio than those with large b/a-ratio. The preferred removal of the b-axis compared to the removal of the a-axis does not apply to these pebbles. Therefore, for these pebbles the b/a-ratio is expected to increase with decreasing $A_{ab}$.

These three areas can be more clearly distinguished in basalt than in calcite because, due to the greater density, the rotation around the longer a-axis is more pronounced. Figure 8 shows the b/a-ratio for 501 elliptical pebbles from basalt as a function of $A_{ab}$. In the range $200 \text{ cm}^2 > A_{ab} > 30 \text{ cm}^2$, the $b/a$-ratio increases with decreasing $A_{ab}$. This is the range of predominant rotation around the long a-axis. Between $30 \text{ cm}^3 > A_{ab} > 6 \text{ cm}^3$ the $b/a$-ratio decreases, because here the rotation around the energetically preferred c-axis dominates. For $A_{ab} < 6 \text{ cm}^2$ the $b/a$-ratio increases again because a stable rotation around the c-axis is no longer guaranteed.

The upper abscissa shows the $b/a$ dependence as a function of the (somewhat more intuitive) mean mass $\langle M \rangle$ of the pebbles. The values of $\langle M \rangle$ were calculated from the area $A_{ab}$ and $c = \lambda \cdot \sqrt{ab}$ : $\langle M \rangle = (4\pi/3)\, \rho_{Bas}\, (ab)^{3/2} \cdot \lambda$. The three ranges $1\,\text{g} \leq \langle M \rangle \leq 20\,\text{g}$ (decrease in $b/a$); $20\,\text{g} \leq \langle M \rangle \leq 250\,\text{g}$ (increase in $b/a$) and $250\,\text{g} \leq \langle M \rangle \leq 2\,\text{kg}$ (decrease in $b/a$) can be distinguished here.

When rotating rapidly around the c-axis, an elliptical pebble makes a transition from rolling to hopping. This occurs if the vertical acceleration of the centre-of-mass satisfies $a_{CM}(\varphi) + g \leq 0$. From fig. 4c, this occurs first for $\varphi = 0$, i.e. at the a-axis. The transition depends on the $b/a$-ratio of the pebble and on the length of the a-axis. The transition from rolling to hopping for a pebble with $a = 2\,\text{cm}$ and $b/a = 0.76$ will occur at the angular velocity

$$\omega_{co} = \sqrt{g/a} \cdot \frac{1}{\sqrt{1-(b/a)^2}} = 34.1\ \text{s}^{-1} \tag{21}$$

which corresponds to a frequency of about $\nu_{co} = 5.42$ Hz.

During water backflow, a pebble will be accelerated until hopping sets in, either at one of the endpoints of the a-axis or in its close vicinity. Thus there is at most half a revolution between the opposite axis endpoints and the point of hopping. The change in angular momentum $\Delta L_\omega = I_c\, \Delta\omega$ is small in this short time, so the hopping distance is also small. Therefore the pebble will hit the beach in the angular range $0 < \varphi \ll \pi/2$. When the pebble hits the beach it experiences a torque $\boldsymbol{T} = \boldsymbol{r}(\varphi) \times \boldsymbol{F} = d\boldsymbol{L}_\omega/dt$ which increases the angular momentum and the angular velocity so that larger hopping widths are then possible. However, if the angle of incident exceeds $\varphi = \pi/2$, the torque $\boldsymbol{T}$ changes sign so that the angular momentum decreases. It is therefore to be expected that if the sandy beach is strongly inclined or if there is a strong backflow of water, a situation will arise in which the pebble hits the ground in the vicinity of the b-axis. This also leads to an effective abrasion of the elliptical pebble around the b-axis.

In sects. 4.2 and 4.3, two erosion processes were discussed, i.e. the abrasion of the a- and b-axis when the pebble rotates around the c-axis and the abrasion of the b- and c-axis during the brief rotation around the a-axis. The abrasion of the c-axis -- when the pebble slides on the sandy beach without a rotation -- depends only on the ratio $b/a$ [13]. The ratios $b/a$ and $c/a$ contain only two of these erosion processes while $c/b$ contains all three processes. It can therefore qualitatively be expected that the half-width of the distributions functions of $b/a$ and $c/a$ are narrower than that for *c/b*. This expectation is confirmed by the empirical result that the half-widths $FWHM_{b/a} \cong FWHM_{c/a} \cong 0.205$ are about 30% smaller compared to the half-width $FWHM_{c/b} = 0.266$ (fig. 2).

## 6. Summary

In this work it was shown that, contrary to intuition, during the energetically preferred rotation of the elliptical pebble around the c-axis, the abrasion of the shorter b-axis is always stronger than that of the longer a-axis. Analogously, the same applies to the brief rotation around the a-axis. Here the abrasion of the c-axis is stronger than that of the b-axis. The *reverse* would be necessary for the formation of spherical pebbles. Therefore, an elliptical pebble on an ocean beach will never become disk-like or spherical. This contradicts Aristotle´s statement [1], at least for elliptical pebbles, that the abrasion process on beaches ultimately leads to *spheres.*

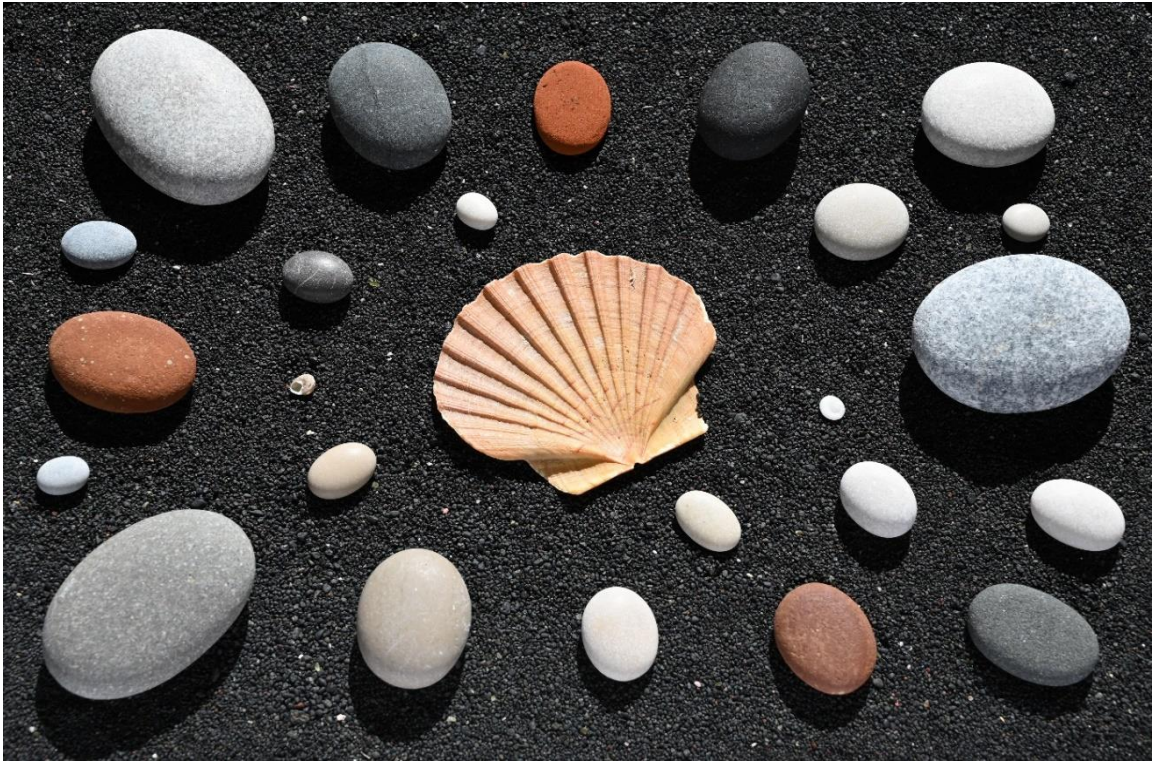

Fig. 1 Twenty-two elliptic pebbles of calcite, basalt, sandstone and brick and a scallop *(pectinidae)* displayed on black sand from the beaches of the island of Lanzarote.

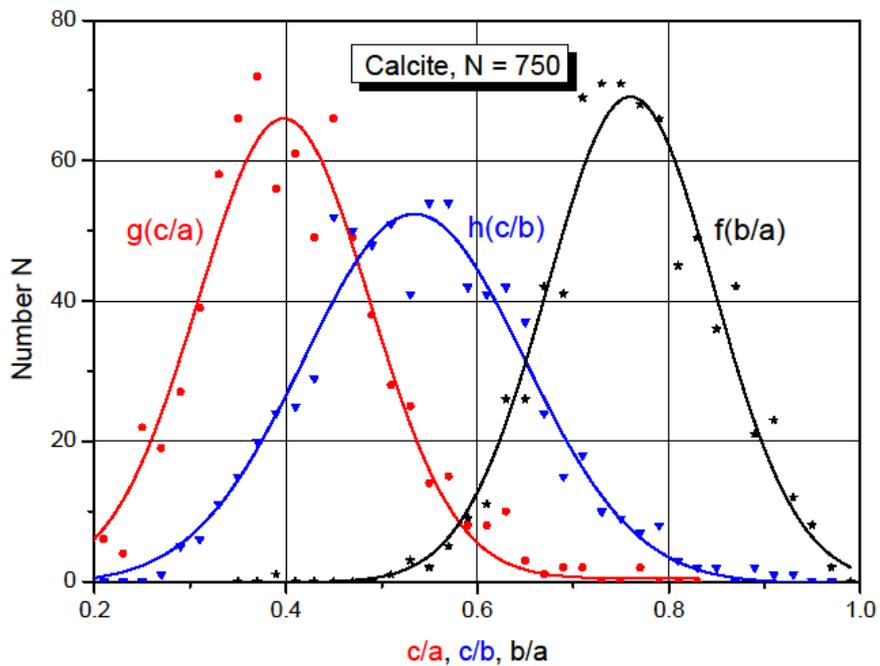

Fig. 2 Frequency distributions for 750 elliptical pebbles of calcite as a function of axes ratios $(b/a)$, $(c/a)$ and $(c/b)$. Plotted is the number $N$ of pebbles in the interval with $\Delta = 0.02$ for all three axial ratios. The data were fitted by Gaussian distributions $f(b/a)$, $g(c/a)$ and $h(c/b)$.

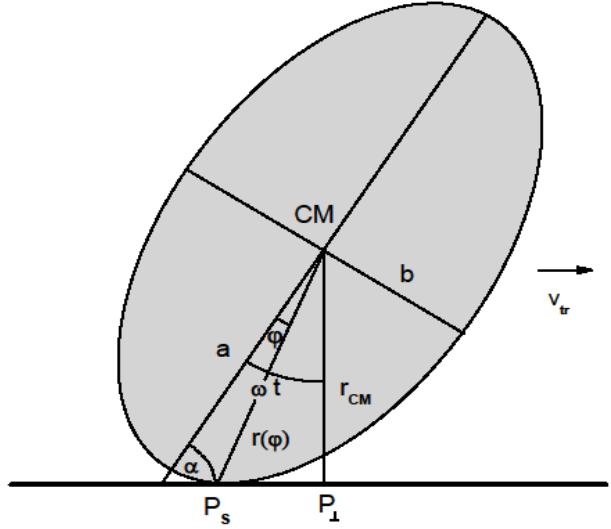

Fig. 3  Rotation of an elliptic pebble around the c-axis, with $\varphi$ the angle between a-axis and $r$ and $(\omega t)$ the angle between a-axis and $r_{CM}$ (schematic).

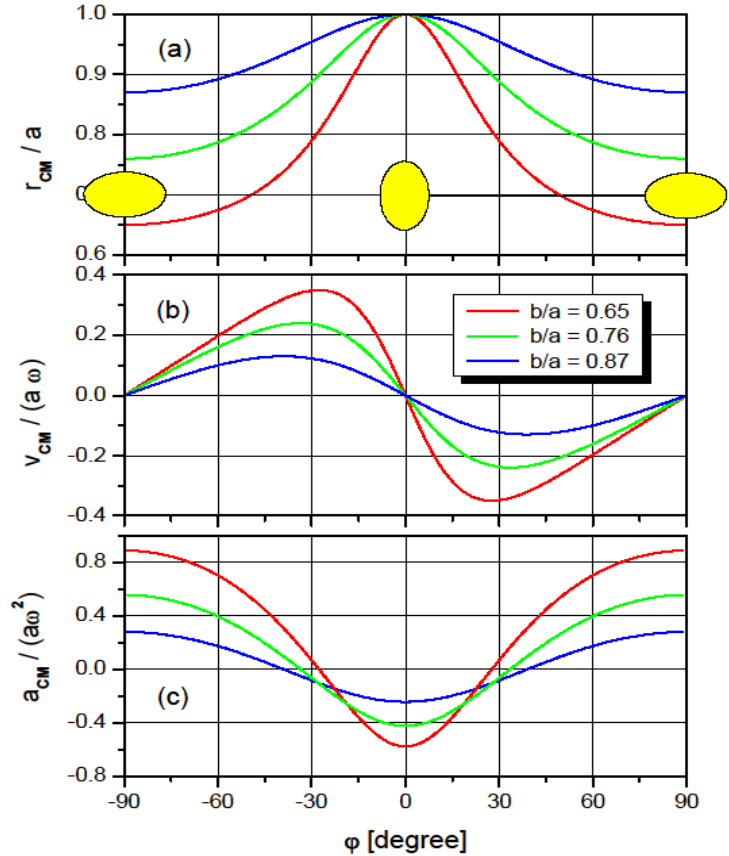

Fig. 4  Centre-of-mass distance $r_{CM}(\varphi)/a$, center-of-mass velocity $v_{CM}(\varphi)/a\,\omega$, and center-of-mass acceleration $a_{CM}(\varphi)/a\,\omega^2$ for three typical $b/a$-ratios as a function of the angle $\varphi$. Note that the latter deviates strongly from that in Ref. [13]. For convenience, at $t=0$ the a-axis direction is chosen vertical.

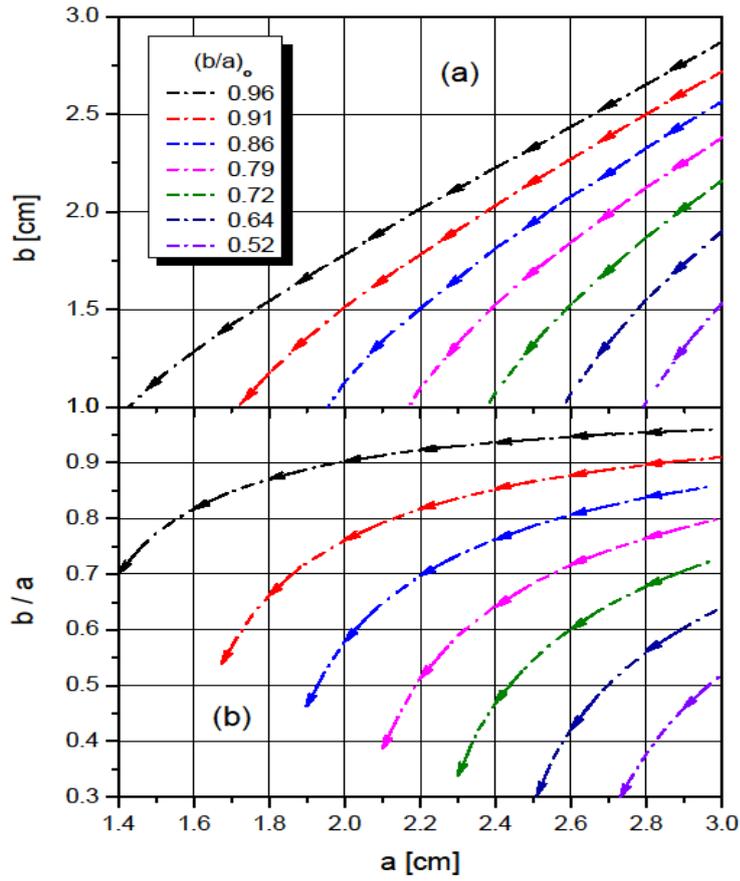

Fig. 5 a) The length of the b-axis as a function of the a-axis for an elliptical pebble with starting value $a_o = 3$ cm and different starting values of $(b/a)_o$.
b) The ratio $b/a$ as a function of the a-axis with the same set of starting values $(b/a)_o$.

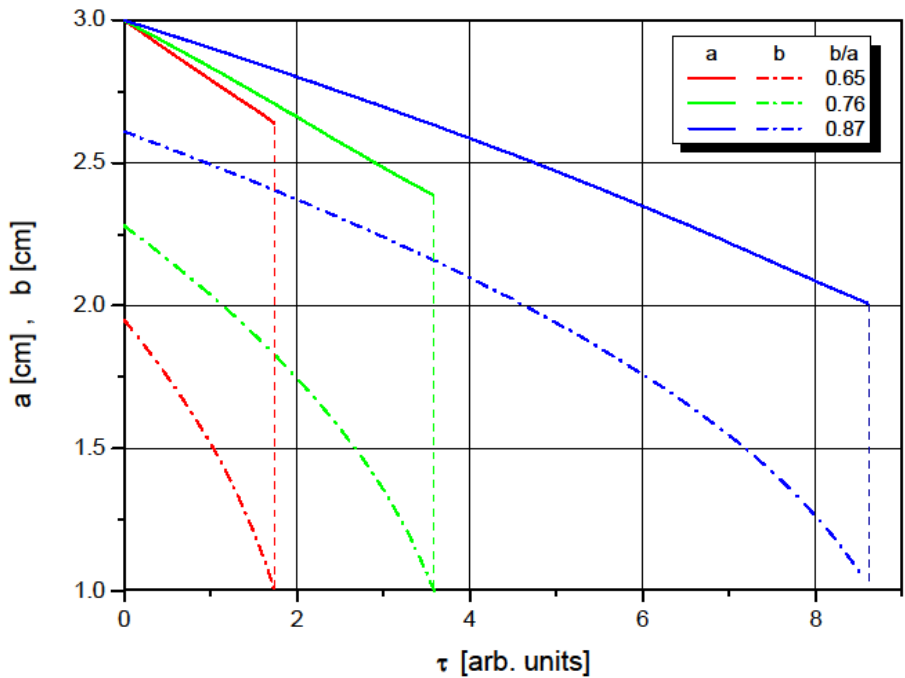

Fig. 6 $a(\tau)$ and $b(\tau)$ for initial value $a(0) = 3$ cm and $(b/a)_o = 0.65;\ 0.76;\ 0.87$.

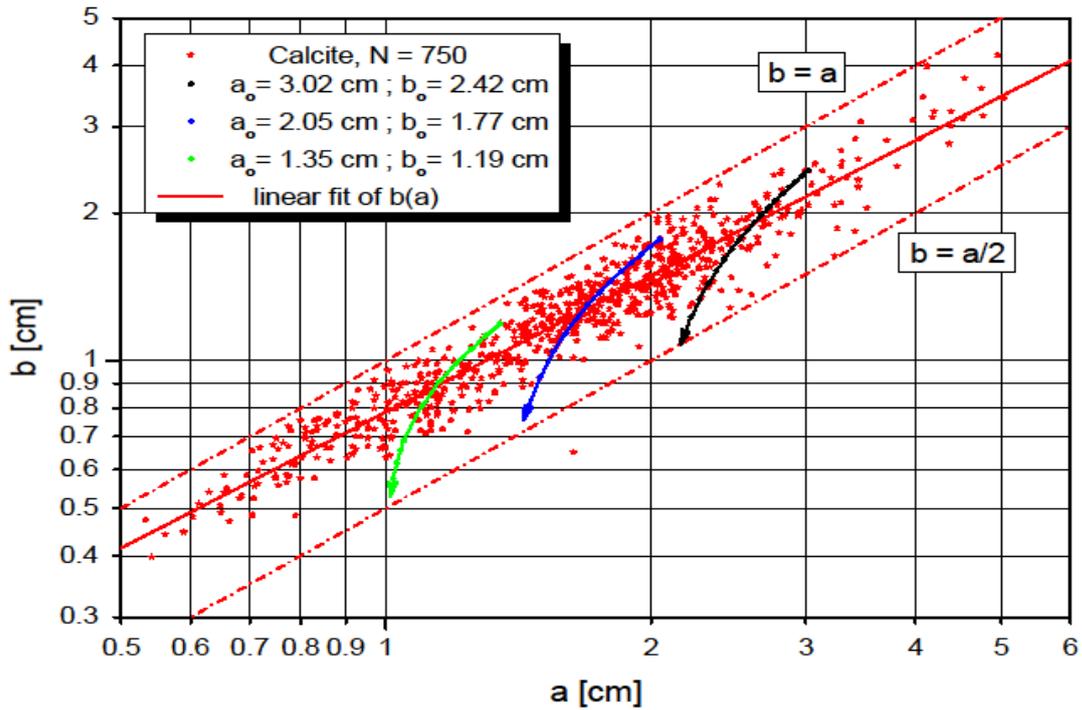

Fig. 7  $b(a)$ for 750 elliptical pebbles of calcite and the time development of $b(\tau)$ for three pebbles with $(b/a)_o = 0.76$ and $a_o = 1.35;\ 2{,}05;\ 3{,}02$ cm.

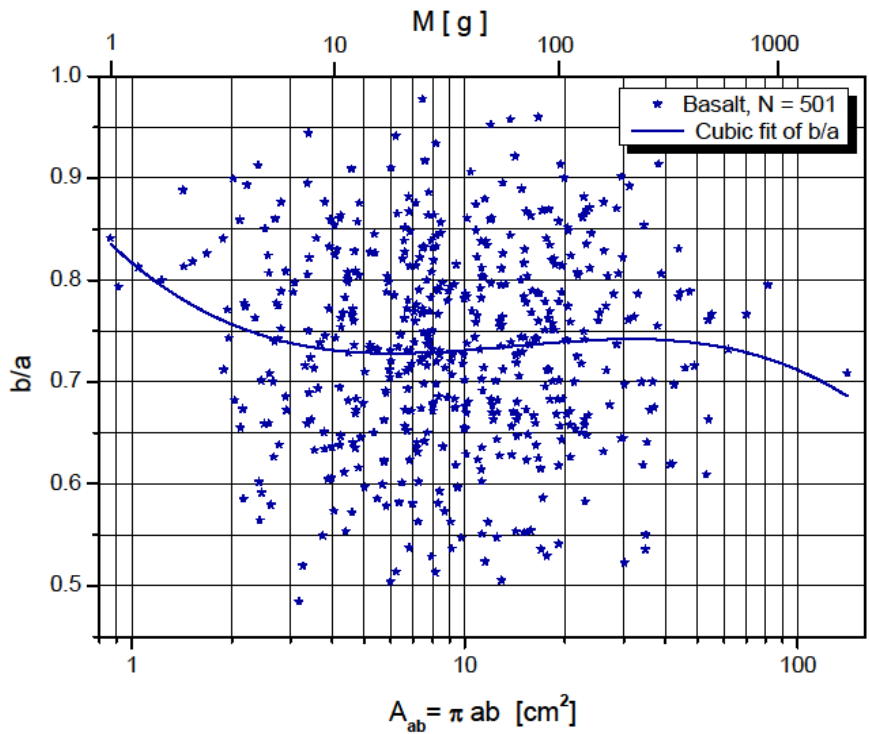

Fig. 8  The $(b/a)$-ratio for 501 elliptical pebbles of basalt as a function of size of the area $A_{ab}$ in the ab-plane with a cubic polynomial fit of the data. On the upper abscissa the calculated masses of the basalt pebbles are given.